# Index Notation of Grid Graphs

Arie Bos [1]

**Abstract:**
*By defining grids as graphs, geometric graphs can be represented in a very concise way.*

## 0 Introduction

In studying fractals like Hilbert curve, Koch snow flake, Sierpinsi triangle and many others [1], the need arose for a universal and compact notation. This paper gives a description. For that a new set of graphs are defined of which the fractals mentioned before are examples. The index notation which goes with it appears to have advantages over existing ones as chain codes and those used in L-systems.

In the next section a rather simple example is given of a graph which will be used to elaborate on the concepts to come. Then in the following two sections the basic concepts of this paper, grid graphs and index notation are introduced.
More examples are given in section 4 where L-systems are introduced, the next section reveals some peculiarities and finally the last section we make an inventory of the fractals which give rise to new integer sequences.

***N.B. 1.*** Apart from the organization of this paper in sections, most of the text consists of statements which are divided in numbered Definitions, Examples, Theorems and Notes, the last ones indicated by N.B. This makes references to text parts much easier.

## 1 Introductory example: Gray codes

**Definition 1.** An ***index notation*** is a bijection $f: S \to T \subseteq \mathbb{Z}$ for some sets *S* and *T*, where $s \in S$ is indicated by $f(s) \in \mathbb{Z}$.

**Example 1.** *S* is the cyclic group of order *n* and $T = \mathbb{Z}/n\mathbb{Z}$, the integers modulo *n*. The elements $s_0, s_1, s_2, s_3, \cdots \in S$ are represented by $0,1,2,3,\cdots$.

***N.B. 2.*** This section is further devoted to Gray codes because these codes are excellent examples to illustrate the new concept of grid graphs and its index notation.

**Definition 2.** A ***binary Gray code*** is an ordered set of binary vectors where each two successive vectors have a (Euclidean or Hamming) distance of 1.

***N.B. 3.*** A Gray code is called ***reflected*** – see Example 2 – if the first half of that code has one particular coordinate equal to 0 and forms a reflected Gray code of 1 dimension less, and the second half is equal to the first half but in opposite order and with that particular coordinate equal to 1.

***N.B. 4.*** A binary Gray code represents a Hamiltonian path on the unit cube.

***N.B. 5.*** There is a lot both known and unknown about Gray codes. One could read interesting facts in [2].

[1] This manuscript has been put on arXiv.org for early dissemination of the results. I welcome any corrections and other comments which could help me to improve this manuscript further; please e-mail me at arie.bos@planet.nl

***N.B. 6.*** The number of Gray codes is enormous [2, p. 13].

**Example 2.** Examples of binary reflected Gray codes in dimensions 0, 1, 2 and 3 are

$$( \ ) \quad \begin{pmatrix} 0 \\ 1 \end{pmatrix} \quad \begin{pmatrix} 0 & 0 \\ 1 & 0 \\ 1 & 1 \\ 0 & 1 \end{pmatrix} \quad \begin{pmatrix} 0 & 0 & 0 \\ 1 & 0 & 0 \\ 1 & 1 & 0 \\ 0 & 1 & 0 \\ 0 & 1 & 1 \\ 1 & 1 & 1 \\ 1 & 0 & 1 \\ 0 & 0 & 1 \end{pmatrix}$$

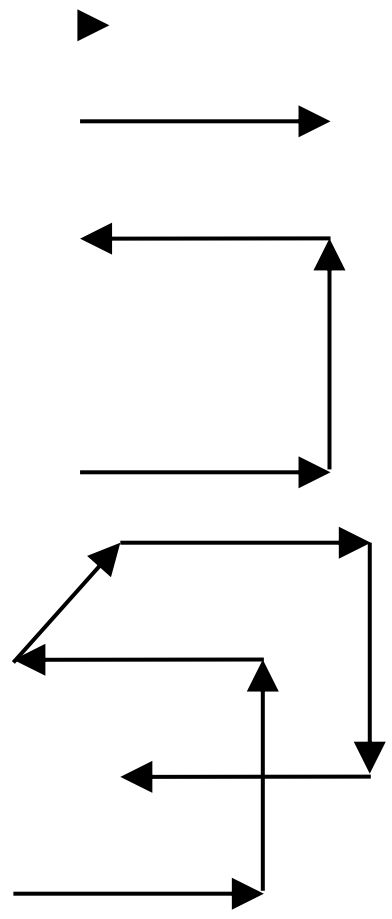

**Example 3.** Binary Gray codes other than reflected ones exist only from dimension 3 on upward. In binary, graph and new notation (which is introduced later in N.B. 8)

$$\begin{pmatrix} 0 & 0 & 0 \\ 1 & 0 & 0 \\ 1 & 1 & 0 \\ 1 & 1 & 1 \\ 1 & 0 & 1 \\ 0 & 0 & 1 \\ 0 & 1 & 1 \\ 0 & 1 & 0 \end{pmatrix}$$

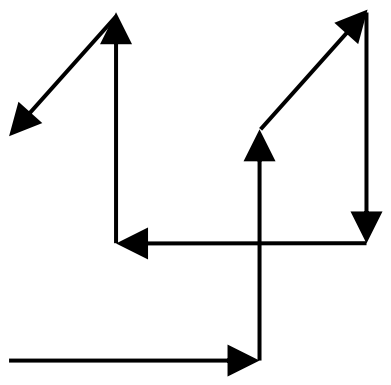

$$\langle 1,2,3,-2,-1,2,-3 \rangle$$

***N.B. 7.*** A binary reflected Gray code will be abbreviated to brGray code.

***Theorem 1.*** There are $d!2^d$ different (but isometric [3]) brGray codes.

**Proof.** The starting point can be chosen in $2^d$ ways and each next point can be chosen in $d!$ ways.

***N.B. 8.*** The coordinates of two successive points $\mathbf{x} = (x_1, \cdots, x_d), \mathbf{y} = (y_1, \cdots, y_d)$ of a binary Gray code differ 1 in only one place: $\begin{cases} x_i = y_i \, , i \neq k \\ |x_k - y_k| = 1 \end{cases}$. We denote this ***edge*** $(\mathbf{x}, \mathbf{y})$ by $\begin{cases} \langle k \rangle \text{ if } x_k + 1 = y_k \\ \langle -k \rangle \text{ if } x_k - 1 = y_k \end{cases}$. Or, in other words, $\langle k \rangle$ if a 1 is added in coordinate *k*, and $\langle -k \rangle$ if a 1 is subtracted in that coordinate. So the Gray code of Example 3 can be denoted by $\langle 1,2,3,-2,-1,2,-3 \rangle$.

***N.B. 9.*** The index notation of a general brGray code becomes $\langle 1,2,-1,3,1,-2,-1,4,1,2,\cdots \rangle$.

## 2 (graph) Grids and grid graphs

***N.B. 10.*** Usually a grid is only a set of (regular) points [4], like $\mathbb{Z}^d$ , or lines [5], but here we make a distinction between vertices and edges, so handling it as a graph. To avoid misunderstanding one can use the term ***graph grid***.

**Definition 3.** A *(graph) grid* $G$ is a tuple of sets of ***vertices*** and ***edges*** as defined below.
Vertices: Let $B=\{\mathbf{b}_1,\cdots,\mathbf{b}_d\}$ be a set of independent vectors[2] of $\mathbb{R}^d$ and

$B^*=\left\{\sum_1^d \beta_i \mathbf{b}_i \mid 0\le \beta_i <1; \beta_i\in\mathbb{R}\right\}$ be the set of points within the block or zonotope [6] with sides $\mathbf{b}_i, i=1,\cdots,d$ . For a (mostly finite) set of points $A\subset B^*$, the ***vertices*** (points) of grid $G$ , determined by $B$ and $A$, are defined by: $V(G)=\left\{\mathbf{a}+\sum_1^d n_i\mathbf{b}_i \mid \mathbf{a}\in A; n_i\in\mathbb{Z}\right\}$, or, loosely notated, $V(G)=A+\mathbb{Z}^d * B$.

Edges: Let $A'=\left\{\mathbf{a}+\sum_1^d n_i\mathbf{b}_i \mid \mathbf{a}\in A; n_i\in\{0,1\}\right\}\supset A$ and $E$ a (finite) set of vectors $E\subset\{\mathbf{x}-\mathbf{y},\mathbf{y}-\mathbf{x} \mid \mathbf{x}\in A',\mathbf{y}\in A\}$, such that $\mathbf{z}\in E\Leftrightarrow -\mathbf{z}\in E$. The ***edges*** of grid $G$ are defined by $E(G)=\left\{\mathbf{e}+\sum_1^d n_i\mathbf{b}_i \mid \mathbf{e}\in E; n_i\in\mathbb{Z}\right\}$. If we have $\mathbf{e}+\mathbf{v}=\mathbf{w}$, for $\mathbf{e}\in E$ and $\mathbf{v},\mathbf{w}\in V(G)$, then $\mathbf{w}-\mathbf{v}$ is an ***edge*** of the grid and it is usually denoted by $(\mathbf{v},\mathbf{w})$.

***N.B. 11.*** $B^*$ is called the ***cell*** of the grid, a ***mesh*** is the smallest infinitely repeated figure, by translation, rotation and/or reflection, (sometimes more than one kind), formed by vertices and edges.

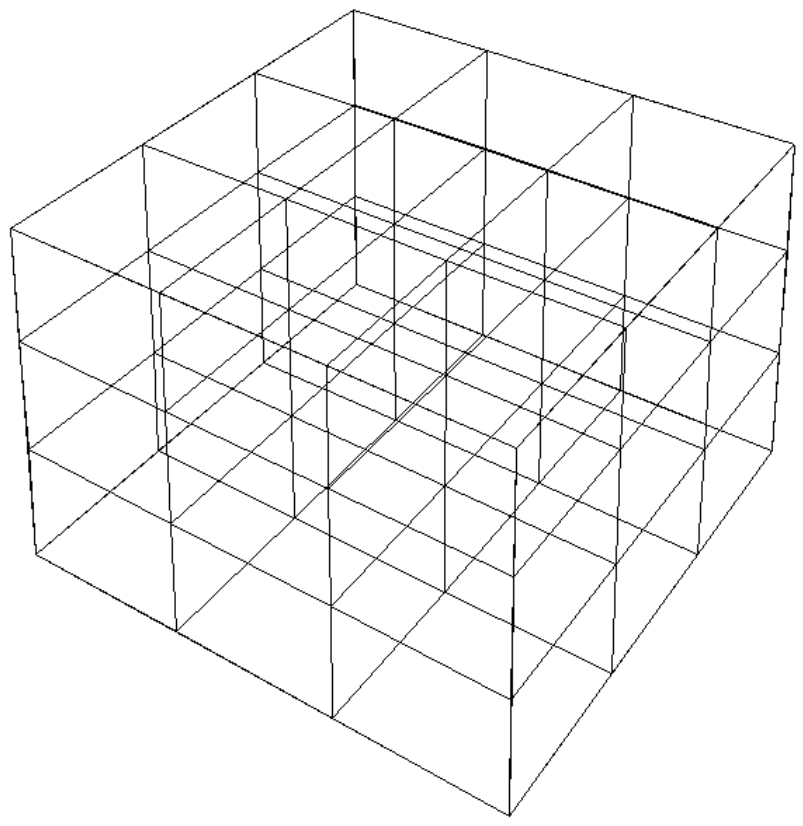

**Example 4.** With $E=B$: let $E=B=\{\pm\mathbf{u}_i \mid i=1,\cdots,d\}$ be the canonical set of unit vectors defined by $\mathbf{u}_i=\left(0,\cdots,0,\underset{j}{1},0,\cdots,0\right)=([i=j])$ (see [7] for the Iverson bracket) and let $A=\{(0,0,\cdots,0)\}$ only contain the origin. Then the ***orthogonal*** grid $G$ has $V(G)=\mathbb{Z}^d$ and the edges are the lines which connect points at mutual distance 1.

***N.B. 12.*** If $E=\{\mathbf{e}_1,\mathbf{e}_2,\cdots,\mathbf{e}_n\}$ is a set of edge vectors, then it is not always sufficient for $(\mathbf{v},\mathbf{w}) \mid \mathbf{v},\mathbf{w}\in V(G)$ to be an edge if $\mathbf{v}-\mathbf{w}\in E$ or $\mathbf{w}-\mathbf{v}\in E$ N.B. 12 shows this.

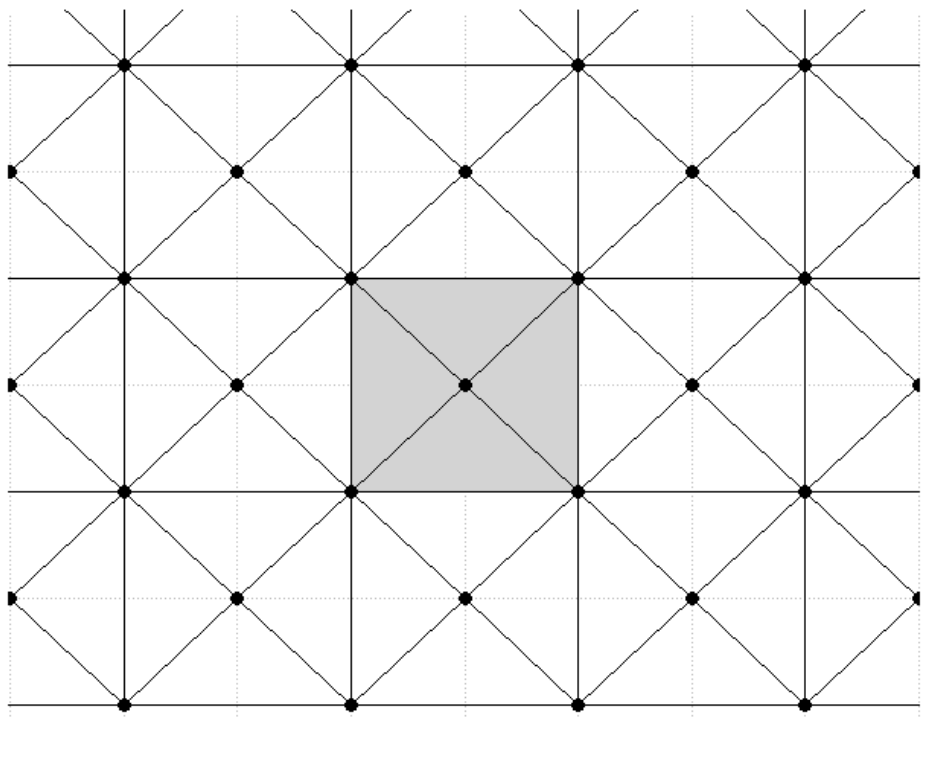

**Example 5.** With $E\supset B$: $d=2$; $B=\{\mathbf{b}_1=(0,2),\mathbf{b}_2=(2,0)\}$; $A=\{(0,0),(1,1)\}$ and $E=\pm B\cup\{\pm(1,1),\pm(-1,1)\}$[3] we get the ***square***

[2] We will make no distinction between a vector and its (end)point. Bot

[3] For a vector $\boldsymbol{x}$ we denote $\{\pm\boldsymbol{x}\}=\{\boldsymbol{x},-\boldsymbol{x}\}$, for a set of vectors $\pm X$

***centered*** grid with the cell $B^*$ in grey. The mesh is the (smallest) rectangular triangle. Notice that $((1,1),(3,1))$ is not an edge, although the difference vector is in *E*.

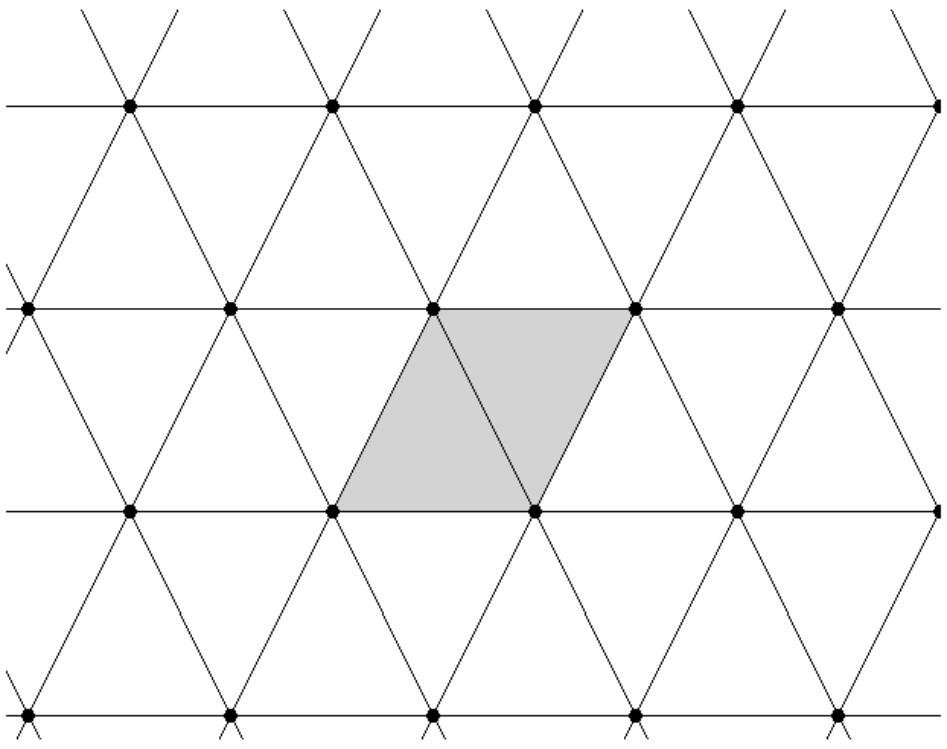

**Example 6.** $d = 2$; $B = \left\{\mathbf{b}_1 = (1,0), \mathbf{b}_2 = \left(\tfrac{1}{2}, \tfrac{1}{2}\sqrt{3}\right)\right\}$; $E = \left\{\pm(1,0), \pm\left(\tfrac{1}{2}, \tfrac{1}{2}\sqrt{3}\right), \pm\left(-\tfrac{1}{2}, \tfrac{1}{2}\sqrt{3}\right)\right\}$ and $A = O = \{(0,0)\}$ only contains the origin. This gives the ***triangular*** grid as is shown, with the cell $B^*$ in grey. The mesh is the equilateral triangle.

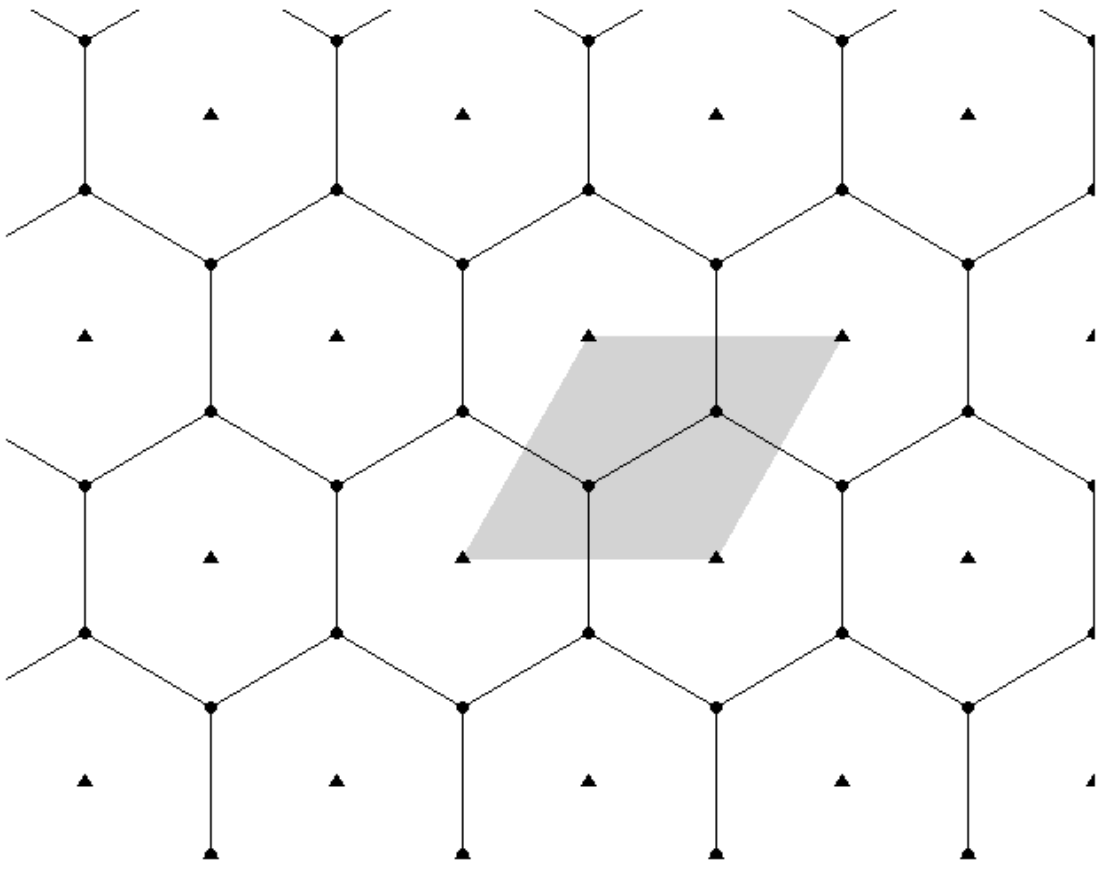

**Example 7.** Similar as Example 6 with $B = \left\{\mathbf{b}_1 = (1,0), \mathbf{b}_2 = \left(\tfrac{1}{2}, \tfrac{1}{2}\sqrt{3}\right)\right\}$, but now with $A = \left\{\left(\tfrac{1}{2}, \tfrac{1}{6}\sqrt{3}\right), \left(1, \tfrac{1}{3}\sqrt{3}\right)\right\}$ and, see figure right, $E = \pm\left\{\left(0, \tfrac{1}{3}\sqrt{3}\right), \left(-\tfrac{1}{2}, \tfrac{1}{6}\sqrt{3}\right), \left(\tfrac{1}{2}, \tfrac{1}{6}\sqrt{3}\right)\right\}$ Vertices of the ***hexagonal*** grid are depicted as dots; the small triangles $\left\{\sum_1^d n_i \mathbf{b}_i \mid n_i \in \mathbb{Z}\right\}$ do not belong to the grid, since $(0,0) \notin A$. Cell $B^*$ in grey again and a hexagon as mesh.

***N.B. 13.*** This last hexagonal grid is the ***dual*** [8] ***grid*** of the triangular one. (A vertex of the dual grid is the center of a mesh of the original grid and the vertices of the dual form an edge if and only if the corresponding meshes share an edge.)

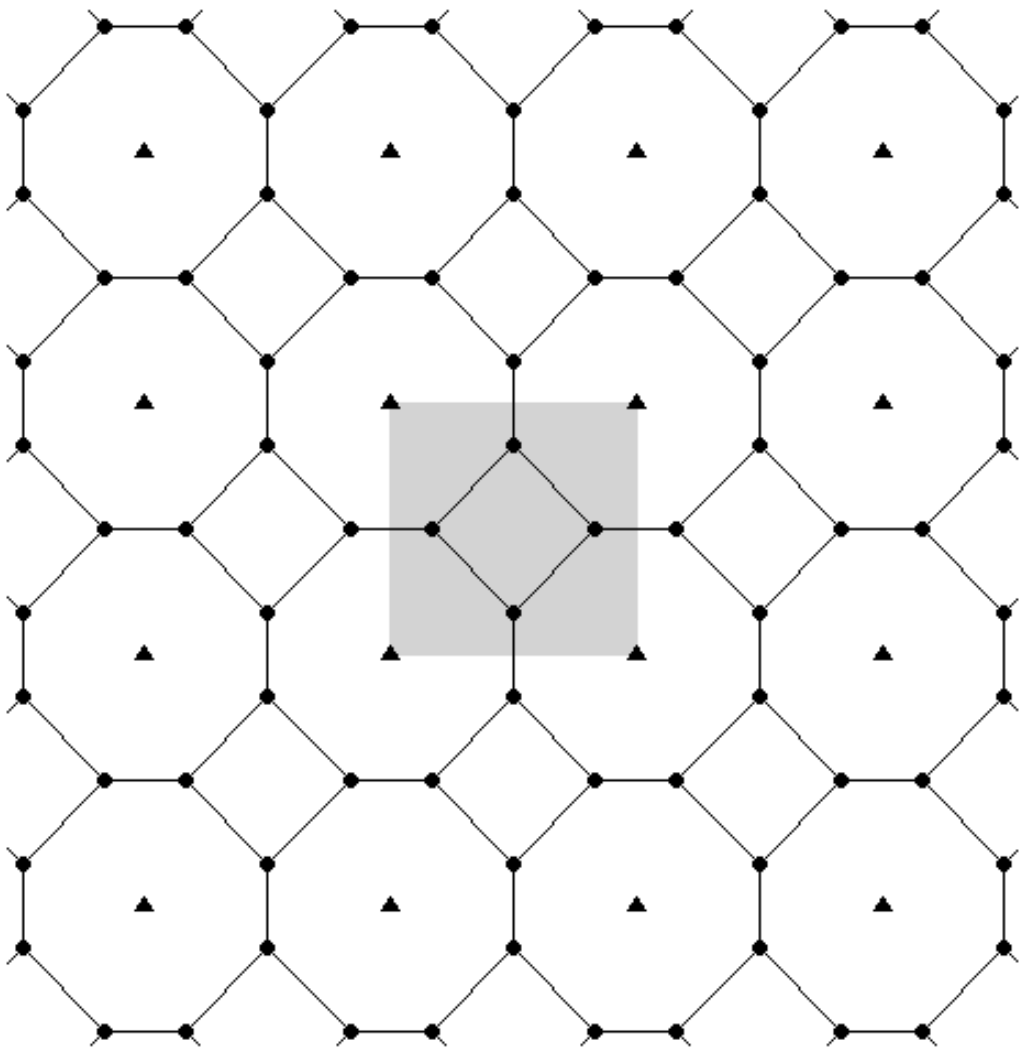

**Example 8.** Next is a grid with (two) different meshes. $B = \{\mathbf{b}_1 = (0,2), \mathbf{b}_2 = (2,0)\}$,

$A = \{(1, 1/3), (1\,2/3, 1), (1, 1\,2/3), (1/3, 1)\}$ and

$E = \pm\{(0, 2/3), (2/3, 0), (2/3, 2/3), (-2/3, 2/3)\}$

This gives two kinds of mesh, a square and a semi regular octagon. Cell $B^*$ in grey again

***N.B. 14.*** From Definition 3 it becomes clear that every grid is isomorphic to a grid defined on $\mathbb{Z}^d$ : a linear map from $\{\mathbf{b}_1, \cdots, \mathbf{b}_d\} \to \{\mathbf{u}_1, \cdots, \mathbf{u}_d\}$ does the job. So without loss of generality, it is sufficient to study these grids, although the pictures can be a bit stretched, as the next example shows.

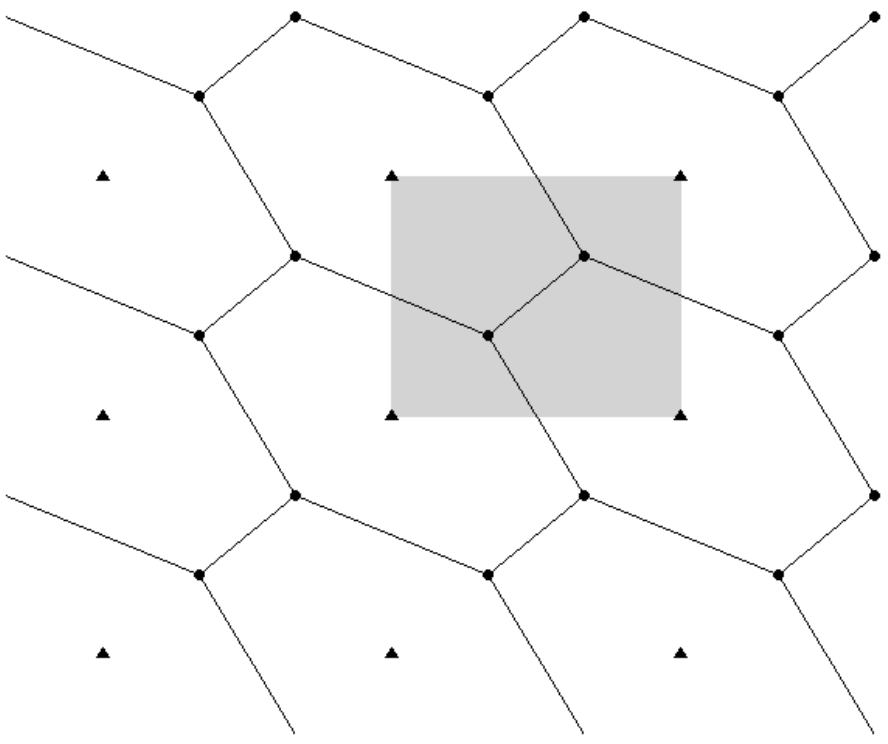

**Example 9.** Analogous to Example 7Example 6, with $B = \{\mathbf{b}_1 = (1,0), \mathbf{b}_2 = (0,1)\}$, $A = \left\{\left(\tfrac{1}{3}, \tfrac{1}{3}\right), \left(\tfrac{2}{3}, \tfrac{2}{3}\right)\right\}$ and $E = \pm\left\{\left(\tfrac{1}{3}, \tfrac{1}{3}\right), \left(\tfrac{1}{3}, -\tfrac{2}{3}\right), \left(-\tfrac{2}{3}, \tfrac{1}{3}\right)\right\}$.

***N.B. 15.*** With Definition 3 it is very easy to project a grid to a lower dimension, only the $B, A$ and $E$ have to be projected, everything else remains the same.

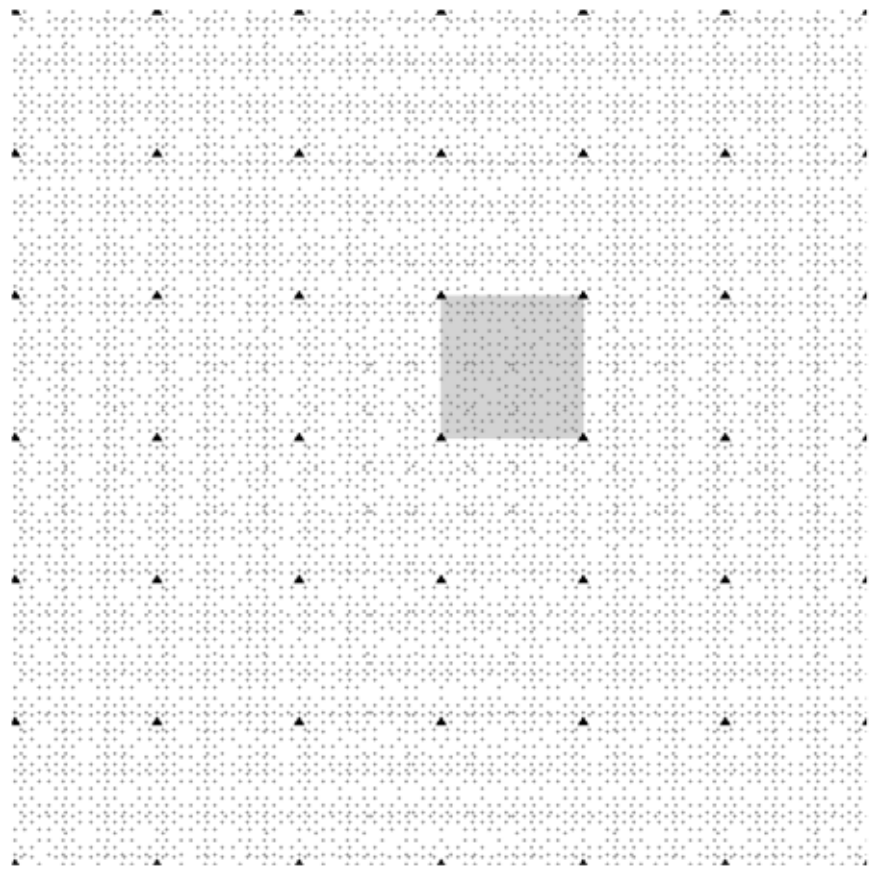

***N.B. 16.*** Rather peculiar grids are the ones generated by the complex *n*$^{th}$ root of unity. Example 6 and Example 7 are connected to the 6$^{th}$ and 3$^{rd}$ root of unity and show a very regular grid. The 8$^{th}$ roots of unity however generate a grid which is dense in $\mathbb{R}^2$ as the figure at the right suggests in which only the vertices are drawn. Most roots of unity behave in this way.

**Definition 4.** A **grid graph** is a subset of a grid, i.e. the vertices are a subset of $V(G)$ and the edges are all the edges in $E(G)$ between the vertices of this subset.

***N.B. 17.*** The term grid graph usually has a far more restricted meaning [9], where it is a rectangular subset of the square graph of N.B. 12.

***N.B. 18.*** All normal concepts of graph theory apply, such as *path*, *tree*, *connected*, etc.

***N.B. 19.*** A lot of fractals 'live' on the grids of Example 4 and Example 6, i.e . they are actual grid graphs, which will be shown later.

**Example 10.** Gray codes live on $\mathbb{Z}^d$ , see Example 2, and curves of Peano, Hilbert and Moore live on $\mathbb{Z}^2$, see next figure in that order.

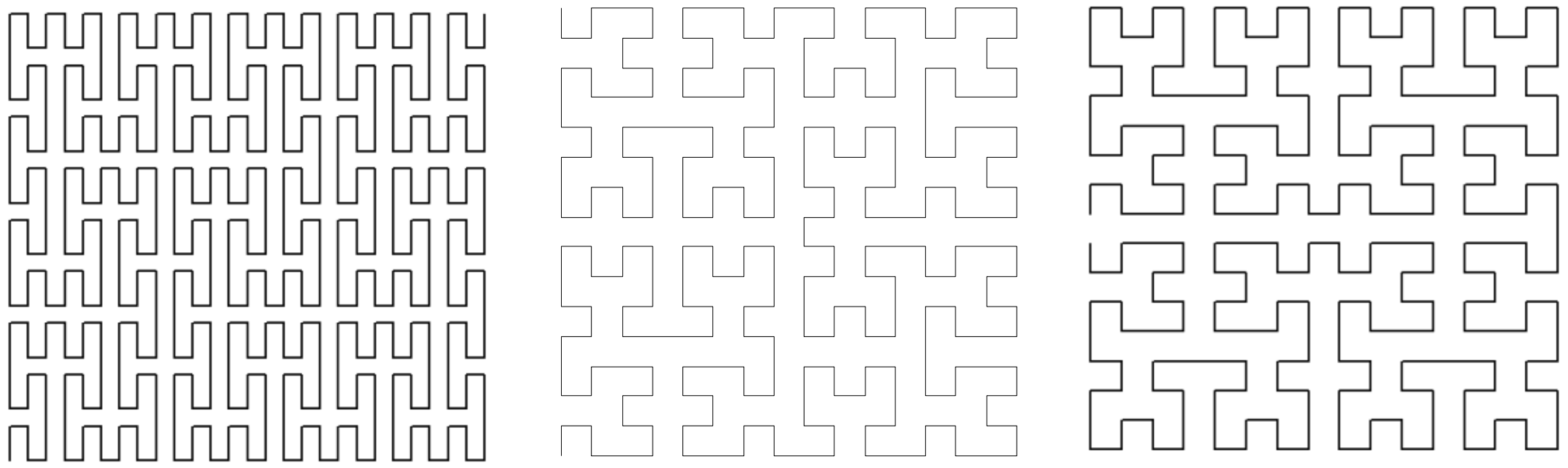

**Example 11.** Koch's snowflake, Sierpinksi's sieve and Gospers flowsnake all live on the triangular grid of Example 6:

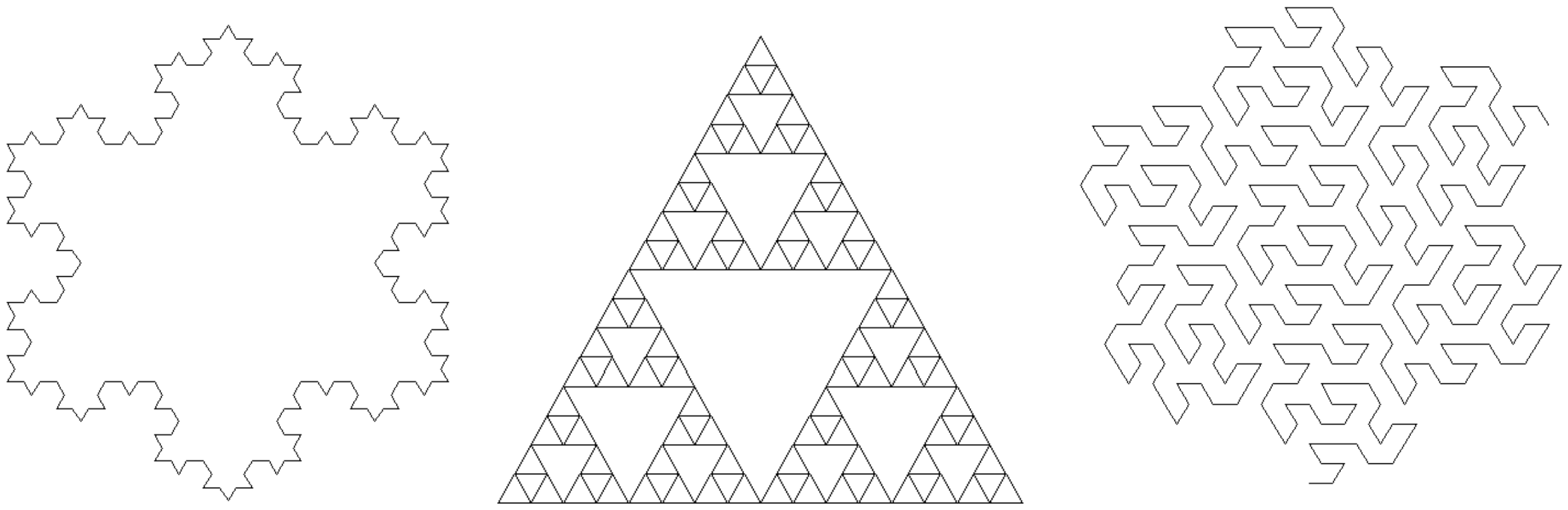

**Definition 5.** A ***walk*** on a grid is a graph consisting of an ordered multiset [4] of vertices such that two subsequent points have an edge in common. See also [10], where ordering is not a prerequisite. A walk is called ***Eulerian*** if each edge in the walk is unique.

**Definition 6.** A ***curve*** is a walk in which not only each edge, but also each vertex is unique.

***N.B. 20.*** So a walk is always a ***directed*** graph.

***N.B. 21.*** Notice that, contrary to a ***path***, a vertex (and even an edge) can be visited more than once in a walk. A path always is a walk.

***N.B. 22.*** Further notice that the edges of a walk form also an ordered set. A walk is determined by its starting point and the subsequent edges. If no starting point is given, the origin is assumed.

***N.B. 23.*** A *random walk* [11] is a good example of a walk. In the next figure an example of a random walk on an hexagonal grid, with probability zero of turning back and equal

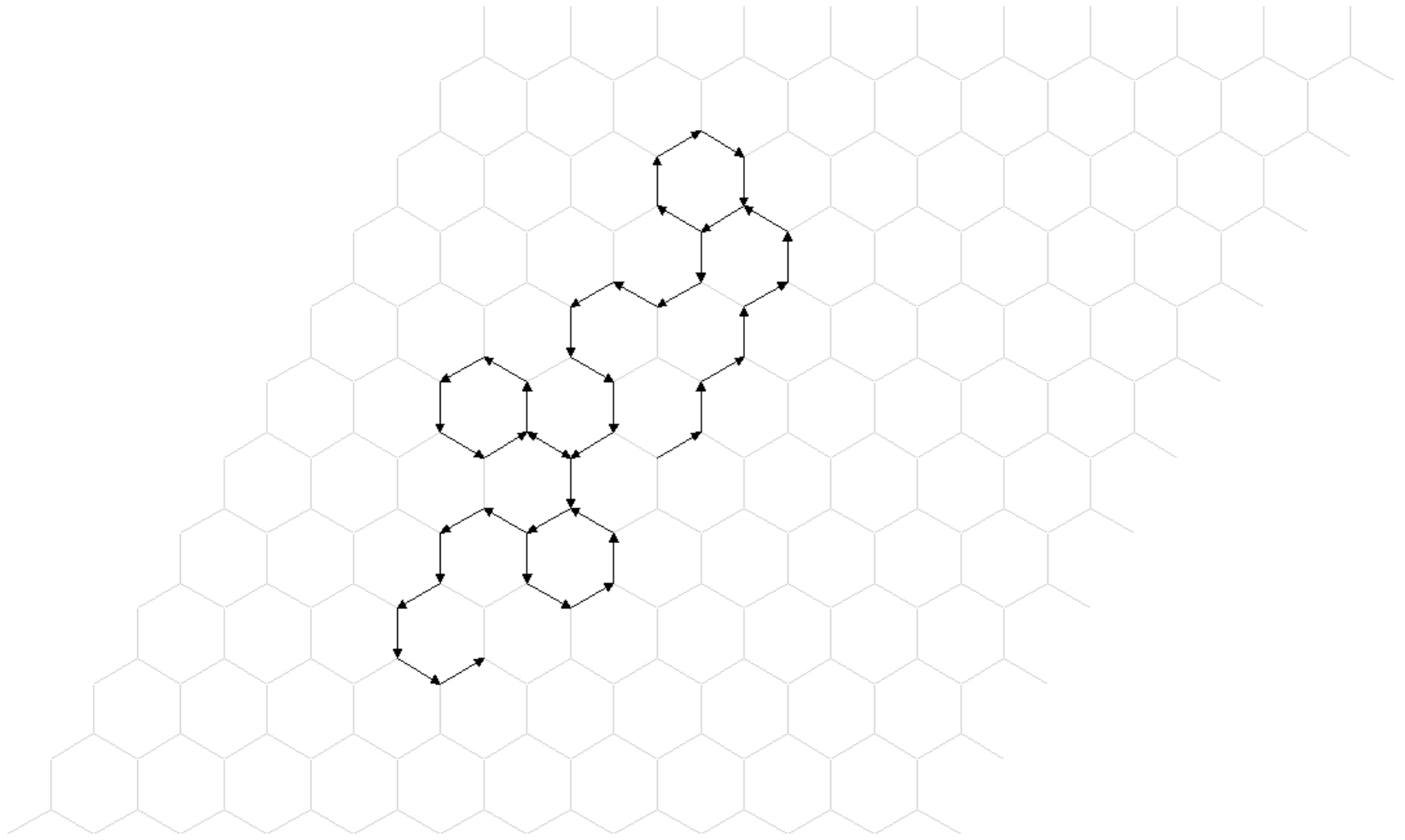

probability of left and right.

***N.B. 24.*** A ***lattice path*** [12] is a walk on $\mathbb{Z}^2$.

***N.B. 25.*** To represented more complicated graphs, walks can be *nested* as follows. If $C_i; i = 1,2,3$ are walks, than $C_1(C_2)C_3$ is the graph where each of the starting points of the two walks $C_2, C_3$ share an edge with the endpoint of $C_1$. In a next paper I will go into several issues on these more general graphs, such as trees.

# 3 Index notation of grid graphs

**Definition 7.** If vertices $\mathbf{x}, \mathbf{y} \in W \subset V(G)$ of a walk $W$ on a grid $G$ are connected and $E = \{\mathbf{e}_1, \mathbf{e}_2, \cdots, \mathbf{e}_n, -\mathbf{e}_1, -\mathbf{e}_2, \cdots, -\mathbf{e}_n\}$, than, by definition, either $(\mathbf{x}, \mathbf{y}) = \mathbf{e}_k \in E \subset E(G)$ or $(\mathbf{y}, \mathbf{x}) = \mathbf{e}_k \in E$ for some *k*. In the first case we denote $(\mathbf{x}, \mathbf{y})$ by $\langle k \rangle$, in the second case by $\langle -k \rangle$. This appears to be an ***indexed edge notation*** or ***indexed edge representation*** of the grid graph, in short ***index notation***. In this way the walk $W$, being an ordered set of edges, can be represented by $W = \langle i, j, k, \cdots \rangle \mid i, j, k, \cdots \in \{1, 2, \cdots, n, -1, -2, \cdots, -n\}$.

[4] An element of a multiset can occur more than once.

***N.B. 26.*** The index notation is very concise. A walk in $\mathbb{R}^d$ consisting of $n+1$ vertices needs $d*(n+1)$ numbers to represent all the coordinates of those ordered points. This we call the ***vertex*** or ***point notation***. For the index notation only the starting point and the edges are necessary, $d+n$ numbers.

***N.B. 27.*** Notice that the index notation is strongly related to the numbering of *E*. Other numbering will give a symmetric grid graph. In a future paper we will go into this.

**Example 12.** The brGray codes in Example 2 can be denoted by $\langle\ \rangle$, $\langle 1\rangle$, $\langle 1,2,-1\rangle$, $\langle 1,2,-1,3,1,-2,-1\rangle$, etc. These codes will be called the ***normalized*** brGray codes. The non reflected Gray code from Example 3 is denoted by $\langle 1,2,3,-2,-1,2,-3\rangle$.

**Example 13.** The *d*-dimensional brGray code is represented by $\langle 1,2,\text{-}1,3,1,\text{-}2,\text{-}1,4,1,2,\text{-}1,\text{-}3,1,\text{-}2,\text{-}1,5,1,2,\text{-}1,3,1,\text{-}2,\text{-}1,\text{-}4,1,2,\text{-}1,\text{-}3,1,\text{-}2,\text{-}1,6,\cdots\rangle$. Here clearly the fractal character can be seen. In a later paper the construction will be shown.

***N.B. 28.*** This last representation was, by the way, in august 2009 a new item in Sloane‘s Encyclopedia of Integer sequences, A164677 [13].

***N.B. 29.*** The index representation of a Gray code is similar to the delta sequence $\delta_n$ of Knuth [2, p. 13]. One can easily see that $\delta_{n-1}+1=|\,e_n|$ for all $n=1,\cdots,2^d-1$. However, it is obvious that the edge sequence determines any walk uniquely (apart from the starting point) and a delta sequence does so only in the case of binary walks, on a cube with side 1.

***N.B. 30.*** Two walks can be concatenated, that is starting the second where the first one ends. Let $C=\langle c_0,c_1,\cdots\rangle$ and $D=\langle d_0,d_1,\cdots\rangle$, then $\langle C,D\rangle=\langle c_0,c_1,\cdots,d_0,d_1,\cdots\rangle$.

***N.B. 31.*** ***Reversing*** a walk is obtained by reversing each of its edges and the order of the edges. So $R^*\langle c_1,\cdots,c_n\rangle=-R\langle c_1,\cdots,c_n\rangle=\langle -c_n,\cdots,-c_1\rangle$, which says that the index representation of a reversed walk $R^*(C)$ is the index representation of *C*, reversed and multiplied by $-1$. $R^*$ will be called the ***negative reverse***, *R* the (normal) ***reverse***.
Notice that $R(R(C))=C$, as you expect, and $R(C_1,C_2)=\langle R(C_2),R(C_1)\rangle$. The same holds for $R^*$. For ease of use we shall omit $R^*$ and only use $-R$.

***N.B. 32.*** If *C* is the index notation of a graph on a particular grid with *n* edges, then this graph can be represented on any grid with *n* edges. Examples of this we will see further on.

***N.B. 33.*** Looking at Example 13 the sequence of edges $\langle 1,-2,-1,4,1,2,-1\rangle$ counts three directions so occupies a 3D space and not a space of smaller dimension. In general, if in the index notation of a (part of a) graph in $\mathbb{R}^d$ only a finite set of numbers is used such that the number of different absolute values equals *m*, then that graphs lives in $\mathbb{R}^m$ and not in a smaller dimension.

# 4 L-systems

***N.B. 34.*** A Lindenmayer system [14], further abbreviated to L-system, is a set {*symbols*, *start*, *(production) rules*}, which produce a set of strings (of variable length). Beginning with the string represented by *start* and applying all the *production rules* to the *symbols* in the successive strings, the succeeding strings are produced. Although *start* usually is one *symbol*, it can also be a string (of *symbols*). We will make this definition a bit more formal. See also [15].

**Definition 8.** An ***L-system*** consists of $V$, an *alphabet* (of symbols), $V^* = V \cup (V \times V) \cup (V \times V \times V) \cup \cdots = \bigcup_{k \geq 1} V^k$ (with * being the Kleene star) the set of all strings (words) of *V*, and ***production rules***, a mapping $P : V \to V^*$.

$P$ can be extended to $P : V^* \to V^*$, where the same letter $P$ is used for the extended mapping, by $P(x_1, x_2, x_3, \cdots) = P(x_1), P(x_2), P(x_3), \cdots$ where $x_i \in V \ (i \geq 1)$ and the strings are concatenated. (Since a concatenation of strings is again a string, the mapping $P$ can be applied $0$ or more times, denoted by $P^n$ for some $n \geq 0$.) $\{x \in V \mid P(x) = x\}$ are ***constants*** and $\{x \in V \mid P(x) \neq x\}$ are ***variables***. For a special $s \in V^*$, mostly $s \in V$, called the ***start***, ***generation*** *n* is $P^n(s)$; the set of all generations is $\{P^n(s) \mid n \geq 0\}$.

***N.B. 35.*** Without loss of generalization we can assume $start \in V$. If *s=start* is a string and not a symbol, we can easily extend *V* with a foreign element (which is not in *V*) , say *t*, and define $P(t) = s$.

**Definition 9.** For $s = (x_1, \cdots, x_n) \in V^*$ the *reverse* $R : V^* \to V^*$ is defined by $R(s) = (x_n, \cdots, x_1)$. Notice that *R* restricted to *V* is the identity.

***N.B. 36.*** A mapping $F : V \to V$ is extended coordinate wise to $F : V^* \to V^*$.

**Example 14.** The ***rabbit sequence*** [16], defined by $V = \{0,1\}$; $P(0) = 1; P(1) = 10$ and $s = 0$. So the generations are

0

1

10

101

10110

10110101

**Example 15.** A slight alteration of the production rules of the rabbit sequence gives the ***Thue-Morse sequence*** [17]: $P(0) = 01; P(1) = 10$ .

***N.B. 37.*** *P* can also be denoted by $\{v \to P(v) \mid v \in V; v \neq P(v)\}$. Or, if $V = \{0, 1, \cdots, m\}$ by $\{P(n) \mid n \leq m\}$ and the start will be $0$, unless stated otherwise.

***N.B. 38.*** To distinguish between strings, they may be enclosed in parentheses and separated by comma's.

***N.B. 39.*** The L-system itself is an abstract mathematical object, an interpretation translates (elements of) this object to graphs or other observable (visual) entities. If we use the index representation of a graph, the pointed brackets $\langle , \rangle$ will be used.

***N.B. 40.*** brGray codes can be generated by the L-system with $V = \mathbb{Z}$, $s = 1$ and rules

$$\begin{cases} P(k) = \langle k+1 \rangle \quad (k > 1) \\ P(1) = \langle 1, 2, -1 \rangle ; P(0) = \langle 0 \rangle \\ P(-k) = -R(P(k)) \ (k < 0) \end{cases} , \text{ where } \langle \pm 1 \rangle = (\pm 1, 0, \cdots) \text{ and } \langle \pm 2 \rangle = (0, \pm 1, 0, \cdots) \text{ etc.}$$

$R$ is the reverse as in N.B. 31. Generation *d* is the brGray code in $\mathbb{Z}^d$ and starts with the string given in Example 13.

**Example 16.** For the Sierpinski triangle from Example 11, we have $V = \{0,1,2,3\}$ and $P = \{\langle 1,2,3\rangle, \langle 1,1\rangle, \langle 2,3,2,1,2\rangle, \langle 3,3\rangle\}$. Here $0$ is the start and $1, 2$ and $3$ represent the three sides of the triangle, given by the vectors $(1,0), \left(-\frac{1}{2},\sqrt{3}\right), \left(-\frac{1}{2},-\sqrt{3}\right)$, which coincide with $E(G)$, the edges of the triangular grid in Example 6, be it in a different order.

**Example 17.** Gosper flowsnake, Example 11, is generated by $V = \{\pm 1, \pm 2, \pm 3\}$; $s = 1$;
$P = \{\langle 1,2,\text{-}1,3,1,1,\text{-}3\rangle, \langle 1,2,2,\text{-}1,\text{-}2,3,2\rangle, \langle 3,1,-3,-2,3,3,2\rangle\}$; $P(-k) = -R(P(k))$ $(k < 0)$
Again $1, 2$ and $3$ represent the three sides of the triangle, represented by the vectors $(1,0), \left(\frac{1}{2},\sqrt{3}\right), \left(-\frac{1}{2},\sqrt{3}\right)$, which determine the triangular grid in Example 6.

A high enough generation appears to be a new integer sequence in [13] and starts with $\langle 1,2,\text{-}1,3,1,1,\text{-}3,1,2,2,\text{-}1,\text{-}2,3,2,3,\text{-}1,\text{-}1,\text{-}3,1,\text{-}2,\text{-}1,3,\text{-}1,\text{-}3,\text{-}2,3,3,2,\cdots\rangle$

***N.B. 41.*** By choosing *V* and *P* appropriately, the index notation of grid graphs often gives rise to new integer sequences in [13]. In section 6 we enumerate a few of them.

***N.B. 42.*** In an L-system repeated squaring can be used easily in the following way. The production rules $P$ are essentially given by $\{P(v) | v \in V\}$. If you apply these rules again on the elements of each $P(v)$, you get $\{P^2(v) | v \in V\}$ The trick now is that this set can be used as new production rules, applied on the elements of each $P^2(v)$, you get $\{P^4(v) | v \in V\}$, which are used as new production rules, etc.

***N.B. 43.*** Applying repeated squaring, the $N = 2^n$ generations are the brGray codes in $\mathbb{Z}^N$

# 5 Some peculiar aspects

***N.B. 44.*** If the string $P(s)$ starts with *s* itself, so $P(s) = (s,S)$ for some string $S \in V^*$, then the generations $P^n(s)$ are ***self-similar*** with $P^{n+1}(s) = P^n(s,S) = (P^n(s), P^n(S))$. Like the accompanying graph, the sequence $\lim_{n\to\infty} P^n(s)$ has a fractal structure. This can be observed in the brGray codes of Example 13 and N.B. 40 and in the rabbit sequence in Example 14.

***N.B. 45.*** In general a self-similar L-system has a recurrent character. That is, $P^n(s)$ can be expressed in terms of $P^k(s); 0 \le k < n$. E.g. in the rabbit sequence of Example 14, we have $P^n(s) = (P^{n-1}(s), P^{n-2}(s))$, which looks very similar to the well-known Fibonacci recurrence.

***N.B. 46.*** Another example of recurrence is our representation of the Sierpinski triangle in Example 16 [5]. If $P^n(0) = \langle F_n, I_n, L_n\rangle$ with $F_n = \underbrace{\langle 1,\cdots,1\rangle}_{2^n\ times}$ and $L_n = \underbrace{\langle 3,\cdots,3\rangle}_{2^n\ times}$, then

[5] Notice that [19] actually does not give triangles.

$P^{n+1}(0) = \langle F_n, P^n(0), I_n, P^n(0), L_n \rangle$. No new sequence (in [13]) however can be derived in this way.

***N.B. 47.*** As pointed out before (in N.B. 32), graphs in the triangular grid can be easily translated to graphs in the hexagonal grid or a graph on $\mathbb{Z}^3$, since all these grids have 3 edges. In the following we see an example.

**Example 18.** Gosper's flowsnake, figure left on a triangular grid, see also Example 11, looks in 3D as in the figure right, seems quite a bit less regular as the left one.

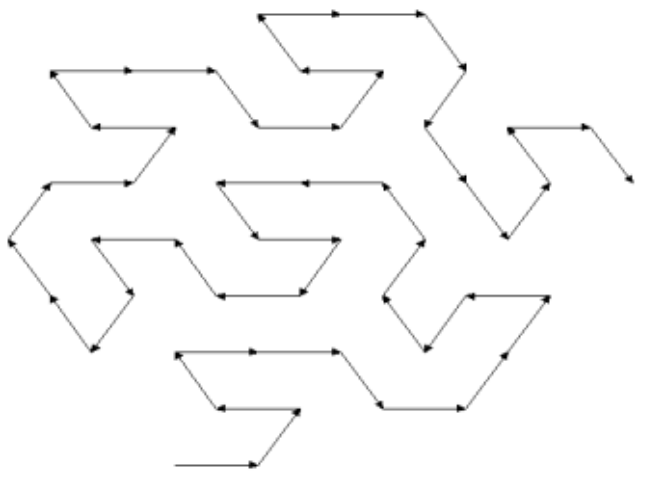

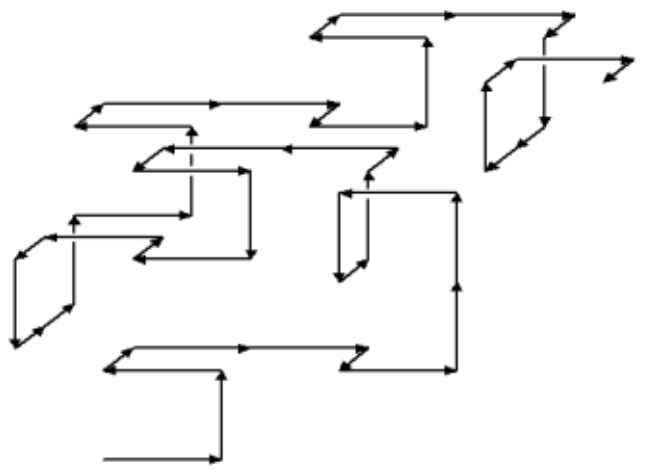

***N.B. 48.*** Some of the grids, like Example 5, Example 7 and Example 8, are peculiar in such a way that a certain edge cannot be followed by an arbitrary other edge. In Example 7 e.g. edge 1 can only be followed by -2 or by 3.

# 6 New integer sequences in [13]

***N.B. 49.*** The first new sequence was that of the brGray code, I submitted it in 2009 and it has id A164677 [13].

***N.B. 50.*** The second surely is Gospers flowsnake Example 17. No match is found in OEIS.

***N.B. 51.*** Another new sequence is reported in a separate paper on 2-dimensional Hilbert curves [18], anticipating this work.

***N.B. 52.*** Probably all sequences representing a Hilbert curve in more than two dimensions also give rise to new integer sequences. In a separate paper some of them will be described.

***N.B. 53.*** It seems worthwhile to investigate other fractals and/or other choices of L-systems producing that fractal so as to generate more new sequences. Unless stated otherwise, in the following examples the first variable mentioned is also the start.

**Example 19.** Koch's curve [19], which lives on $\mathbb{Z}^2$, is given by $V = \{\pm 1, \pm 2\}$ and $P(1) = \langle 1,2,1,-2,1 \rangle$; $P(-1) = \langle -1,-2,-1,2,-1 \rangle$; $P(2) = \langle 2,-1,2,1,2 \rangle$; $P(-2) = \langle -2,1,-2,-1,-2 \rangle$ then we get the sequence 1,2,1,-2,1,2,-1,2,1,2,1,2,1,-2,1,-2,1,-2,-1,-2,1,2,1,-2,1,2,-1, 2,1,2,-1,-2,-1,2,-1,2,-1,2,1,2,1,2,1,-2,1,2,-1,2,1,2,1,2,1,-2,1,2,-1,2,1,2,1,2,1,-2,1,….. which is new to the OEIS [13].

**Example 20.** An alternative construction for the Sierpinski triangle in [20] is given by the ***Sierpinski arrowhead curve*** [21]. Its production rules and alphabet are $P(1;2;3;-3;-2;-1) = \langle 1,-3,2 \rangle; \langle 3,-2,1 \rangle; \langle 2,-1,3 \rangle; \langle -3,1,-2 \rangle; \langle -1,2,-3 \rangle; \langle -2,3,-1 \rangle$ and $V = \{\pm 1, \pm 2, \pm 3\}$. This produces the sequence 1,-3,2,-3,1,-2,3,-2,1,-3,1,-2,1,-3,2,-1,2,-3,2,-1,3,-1,2,-3,1,-3,2,-3,1,-2,1,-3,2,-1,2,-3,1,-3,2,-3,1,-2,3,-2,1,-2,3,-1,3,-2,1,-3,1,-2,3,-2,1,-2,3… which also is unknown to OEIS [13].

**Example 21.** The ***Dragon curve*** [22] is also on $\mathbb{Z}^2$ and is given by $V = \{\pm 1, \pm 2\}$ and $P(1) = \langle 1,-2 \rangle; P(-1) = \langle -1,2 \rangle; P(2) = \langle 1,2 \rangle; P(-2) = \langle -1,-2 \rangle$ The also new in OEIS sequence is 1,-2,-1,-2,-1,2,-1,-2,-1,2,1,2,-1,2,-1,-2,-1,2,1,2,1,-2,1,2,-1,2,1,2,-1,2,-1,-2,-1,2,1,2,1,2, 1,2,1,-2,-1,-2,1,-2,1,2,-1,2,1,2,1,-2,1,2,-1,2,1,2,-1,2,-1,-2,-1,2,1,2,1,-2,1,2,1,-2,-1,…

$$P(0;1;2;3;-3;-2;-1) = \langle 1,2,3,-1,-2,-3\rangle ; \langle 1,-3,1\rangle ; \langle 2,1,2\rangle ; \langle 3,2,3\rangle ; \langle -3,-2,-3\rangle ; \langle -2,-1,-2\rangle ; \langle -1,-3,-1\rangle$$

**Example 22.** ***Gosper island*** [23] has alphabet $V = \{0, \pm 1, \pm 2, \pm 3\}$ and production rules which leads to the integer sequence 1,-3,1,-3,-2,-3,1,-3,1,-3,-2,-3,-2,-1,-2,-3,-2,-3,1,-3,1,-3,-2,-3,1,-3,1,2,1,2,1,… again unknown to OEIS [13].

**Example 23.** New also is the ***Lévy C curve*** [24] with the same alphabet as the Dragon curve, but slightly different rules: $P(1) = \langle 1,2\rangle ; P(2) = \langle 2,-1\rangle$ and $P(-x) = -P(x)$. The generated sequence is 1,2,2,-1,2,-1,-1,-2,2,-1,-1,-2,-1,-2,-2,1,2,-1,-1,-2,-1,-2,-2,1,-1,-2,-2,1,-2,1,1,2,2,-1,-1,… new in OEIS [13].

**Example 24.** The final example is the ***Takagi curve*** (or ***blancmange curve***) as shown in [25]. Here the alphabet is $\mathbb{Z}$ , the start is *0*, and the production rule is a formula: $P(n) = \langle n+1, n-1\rangle \quad (n \in \mathbb{Z})$. $P^6(0)$ gives the sequence 6,4,4,2,4,2,2,0,4,2,2,0,2,0,0,-2,4,2,2,0,2,0,0,-2,2,0,0,-2,0,-2,-2,-4,4,2,2,0,2,0,0,-2,2,0,0,-2,0,-2,-2,-4,2,0,0,-2,0,-2,-2,… with the half of the difference of two consecutive numbers equal to 1,0,1,-1,1,0,1,-2,1,0,1,-1,1,0,1,-3,1,0,1,-1,1,0,1,-2,1,0,1,-1,1,0,1,-4,1,0,1,-1,1,0,1,-2,1,0,1,-1,1,0,1,-3,1,0,1,-1,1,0,1,-2,1,0,1,… which is A088705 in [13].

# 7 Concluding

***N.B. 54.*** In a future paper it will appear to be fruitful to combine index notation with isometric transformations. For that the concept of L-systems is applied in a new way.

***N.B. 55.*** Since we then have a rather complete picture of index notation, it is worthwhile to compare it with other ways of representing curves, like chain codes or turtle graphics.

# 8 Bibliography

[1] Wolfram_f, „Fractals,” [Online]. Available: http://mathworld.wolfram.com/Fractal.html.

[2] D. E. Knuth, "The Art of Computer Programming, volume 4, fascicle 2," Addison Wesley, 2005, p. 13.

[3] „Isometry,” [Online]. Available: http://en.wikipedia.org/wiki/Isometry.

[4] Wolfram_l, „Lattice,” [Online]. Available: http://mathworld.wolfram.com/PointLattice.html.

[5] Wolfram_g, „Grid,” [Online]. Available: http://mathworld.wolfram.com/Grid.html.

[6] Wolfram_zt, „Zonotope,” [Online]. Available: http://mathworld.wolfram.com/Zonotope.html.

[7] D. Knuth, "Two Notes on Notation," Vols. 99, Number 5, May 1992, pp. 403–422, no. American Mathematical Monthly.

[8] Wolfram. [Online]. Available: http://mathworld.wolfram.com/DualTessellation.html.

[9] Wolfram_gg, "Grid Graph," [Online]. Available: http://mathworld.wolfram.com/GridGraph.html.

[10] Wolfram_w, „Walk,” [Online]. Available: http://mathworld.wolfram.com/Walk.html.

[11] Wolfram_rw, "Random walk," [Online]. Available: http://mathworld.wolfram.com/RandomWalk.html.

[12] Wolfram_LP. [Online]. Available: http://mathworld.wolfram.com/LatticePath.html.

[13] N. Sloane, „Integer sequnces,” 2009. [Online]. Available: http://www.research.att.com/~njas/sequences/A164677.

[14] P. a. L. A. Prusinkiewicz, The Algorithmic Beauty of Plants, New Yor: Springer-Verlag, 1990.

[15] String_substitution. [Online]. Available: http://en.wikipedia.org/wiki/String_substitution#String_substitution.

[16] Wolfram_rs, "Rabbit sequence," [Online]. Available: http://mathworld.wolfram.com/RabbitSequence.html.

[17] Thue-Morse. [Online]. Available: http://en.wikipedia.org/wiki/Thue%E2%80%93Morse_sequence.

[18] A. Bos, *Hilbert curves in 2 dimensions generated by L-systems (preprint).*

[19] Koch. [Online]. Available: http://en.wikipedia.org/wiki/L-system#Example_4:_Koch_curve.

[20] Sierpinski. [Online]. Available: http://en.wikipedia.org/wiki/Lindenmayer_system#Example_6:_Sierpinski_triangle.

[21] Sierpinski_arrowhead_curve. [Online]. Available: http://en.wikipedia.org/wiki/Sierpi%C5%84ski_arrowhead_curve.

[22] Dragon_curve. [Online]. Available: http://en.wikipedia.org/wiki/Dragon_curve.

[23] Gosper_Island. [Online]. Available: http://mathworld.wolfram.com/GosperIsland.html.

[24] Lévy_C_curve. [Online]. Available: http://en.wikipedia.org/wiki/L%C3%A9vy_C_curve.

[25] „Tagaki Wolfram,” [Online]. Available: http://demonstrations.wolfram.com/TakagiCurve/.